\documentclass[twocolumn,showpacs]{revtex4}
\usepackage{graphicx}
\usepackage{dcolumn}

\begin{document}


\title{Effects of resonant single-particle states on pairing correlations}

\author{Munetake Hasegawa$^{1}$}
\author{Kazunari Kaneko$^{2}$}
\affiliation{
$^{1}$Laboratory of Physics, Fukuoka Dental College, Fukuoka 814-0193, Japan \\
$^{2}$Department of Physics, Kyushu Sangyo University, Fukuoka 813-8503, Japan
}

\date{\today}

\begin{abstract}

  Effects of resonant single-particle (s.p.) states on the pairing correlations
 are investigated by an exact treatment of the pairing Hamiltonian on the Gamow
 shell model basis.  We introduce the s.p. states with complex energies into
 the Richardson equations.  The solution shows the property that the resonant
 s.p. states with large widths are less occupied.
 The importance of many-body correlations between bound and resonant prticle pairs 
 is shown. 

\end{abstract}


\maketitle

  One of current topics is concerned with the microscopic structure of nuclei
 far from the $\beta$-stability line.  In such nuclei, weak binding of nucleons
 is expected to cause exotic situations where resonant and continuum single-particle
 (s.p.) states contribute considerably to nuclear properties even at low energy.
 The problem has been considered since an early time \cite{Migdal}, and recent
 experimental study of drip-line nuclei has stimulated a new interest in it.
  The resonant and continuum s.p. states could enhance pairing correlations
 \cite{Bennett,Dobac} which are most important in nuclei once the Hartree-Fock (HF)
 or shell-model mean field is formed. They could affect binding energies
 of bound states and nuclear resonant states, and could play important roles
 in properties of these low-energy states.
  Recently, theoretical studies \cite{Krup,Sand,Betan,Michel} on the contributions
 from the resonant and continuum states have been reported.  Refs. \cite{Krup,Sand}
 incorporated the effect of the resonant continuum or resonant states
 in the HF+BCS approximation.  Refs. \cite{Betan,Michel} studied the effects
 of the resonant and continuum s.p. states on one pair state outside a core.

  The purpose of this paper is to investigate the effects of the resonant
 s.p. states on many-body correlations in many particle systems
 (including correlations between bound and resonant particle pairs),
 on the shell-model basis.  We treat the problem by exactly solving the pairing
 Hamiltonian including resonant s.p. states.  Namely, we extend the Richardson
 equations \cite{Rich1} so as to include s.p. states of complex energies and
 solve them.
 The Richardson equations have a defect that the pairing interaction strength
 cannot be changed for bound and resonant s.p. states.
 Still, this approach is very useful for getting a general perspective of our subject.

   We consider a system of Fermions in which the mean field and residual
 interaction are approximately represented by a one-body potential and the
 pairing force. We denote a Fermion by $c^\dagger_{\nu \sigma}$ ($\sigma=\pm$
 denote time-reversal conjugate states) and a pair operator by
 $S^\dagger_\nu=c^\dagger_{\nu +}c^\dagger_{\nu -}$.
 The eigenstates of our system are classified by the number of unpaired Fermions.
 The states with no unpaired Fermion are lower in energy
 than the others. We pay our attention only to such pair-correlated states
 composed of the basis states
 ${S^\dagger_\mu S^\dagger_\nu \cdots} |0 \rangle$.  In this subspace,
 the pairing Hamiltonian is written as 
 ${H=\sum_\nu 2\varepsilon_\nu N_\nu - G \sum_{\mu \nu} S^\dagger_\mu S_\nu}$,
 where $\varepsilon_\nu$ denotes the s.p. energies and $G$ is the pairing force
 strength.
 The operator ${N_\nu = S^\dagger_\nu S_\nu}$ satisfies the relation
 ${[N_\nu , S^\dagger_\nu ]=S^\dagger_\nu}$ from the Pauli principle
 ${c^\dagger_{\nu \sigma} c^\dagger_{\nu \sigma} =0}$, and counts
 the pair number in the level $\nu$.
 Richardson \cite{Rich1} proved that a pair-correlated eigenstate is expressed as
$ \prod^N_{k=1} \{ \sum_\nu {S^\dagger_\nu / ({2\varepsilon_\nu - z_k})} \}
   |0\rangle $, 
 using the roots $z_k$ of the coupled equations
\begin{equation}
  \sum_\nu {1 \over {2\varepsilon_\nu - z_k}} = {1 \over G}+
    \sum_{k^\prime(\ne k)} {2 \over {z_{k^\prime} - z_k}}.    \label{eq:2}
\end{equation}
 Here, the number of the equations (\ref{eq:2}) is the number of pairs $N$.
 For an eigenstate, the total energy and occupation probability of the s.p. level
 $\nu$ are given by
\begin{eqnarray}
 & & E= \langle H \rangle = \sum^N_{k=1} z_k,      \label{eq:3} \\
 & {} & v^2_\nu = \langle N_\nu \rangle =
   -G^2 { d \over dG } \sum^N_{k=1} {1 \over {2\varepsilon_\nu - z_k}}.
    \label{eq:4}
\end{eqnarray}

  Although Richardson \cite{Rich1} supposed $\varepsilon_\nu$ to be real number,
 his proof is not relevant to whether $\varepsilon_\nu$ are real or complex.
  We can therefore introduce resonant s.p. states with complex energies to our
 pairing problem by considering a Woods-Saxon potential. 
 In compensation for this, however, the Hamiltonian stops being Hermitian.
 Even when the s.p. energies $\varepsilon_\nu$ are real, some of the roots $z_k$
 are complex conjugate (the total energy of the system is real). 
  If any of $\varepsilon_\nu$ are complex in Eqs. (\ref{eq:2}),
 all the roots $z_k$ are complex but not complex conjugate.
 If the resonant s.p. states participate in the pairing correlations, therefore,
 the total energies of the pair-correlated states including bound states become
 complex in general. 

   Let us first investigate basic properties of a general system 
 which has resonant s.p. states. We consider a model that has a dozen
 s.p. levels with equal spacing $e$, where the lower eight levels are bound states
 and the upper four levels have the same resonant width $\gamma$,
\begin{equation}
  \varepsilon_\nu = \left\{
    \begin{array}{ll}
      (\nu - 8.5)e              & \mbox{for } \nu=1,2, \cdots 8  \\
      (\nu - 8.5)e - i (\gamma/2) e  & \mbox{for } \nu=9,10,11,12 .
    \end{array} \right.   \label{eq:5}
\end{equation}
We use the energy unit $e=1$ (also for $G$ and $\gamma$) in the following pages.

  When the pair number $N$ is small, the Fermi surface lies deeply below the
 resonant s.p. levels. The resonant states can be neglected as
 long as the pairing force is not very strong. This is the case of
 ordinary treatment for the nuclear pairing problem. 
 Calculated results for the $N=4$ system are summarized as follows.
 The system gets additional energy gain when the resonant levels
 are taken into account and the energy gain increases more rapidly than that
 in case of no resonant levels as the pairing force becomes stronger. 
 The total energy becomes complex, namely, eigenstates including the ground state
 have finite resonant width when the resonant s.p. states contribute to
 the pairing correlations, as mentioned above. Interestingly, the ground state
 has pair decay width larger than low-energy excited states.

\begin{figure}[b]
\includegraphics[width=6.8cm,height=6.6cm]{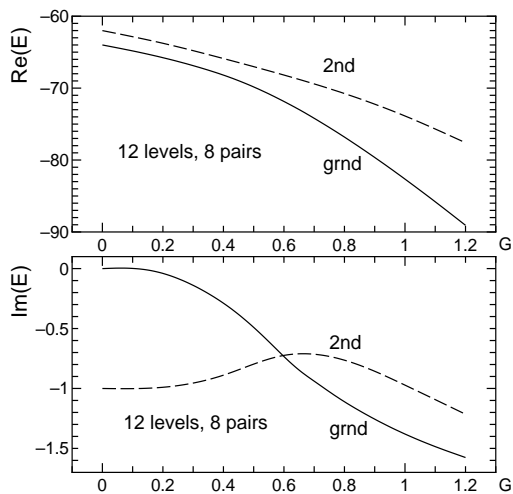}
  \caption{Energies of the ground and next pair-correlated states of the $N=8$ system
           when $\gamma/2 = 0.5$.  Variations of real and imaginary parts depending
           on $G$ are illustrated.}
  \label{fig1}
\end{figure}

  We are interested in the situation of weak binding of nuclei where 
 the Fermi surface lies near the resonant s.p. levels.  Fig. 1 shows the effects
 of the resonant s.p. levels on the $N=8$ system with $\gamma/2=0.5$,
 in which the Fermi surface is between the bound and resonant levels.
 If we do not take account of the resonant levels, there is
 no configuration mixing like the closed shell ($v^2_\nu=1$ for $\nu \le 8$) and
 hence the energy gain due to the pairing correlations increases monotonously
 depending on $G$.  Fig. 1 indicates more rapid increase of the energy gain
 when the resonant s.p. states are taken into account.
 The imaginary part of the total energy shown in the lower graph of Fig. 1
 displays interesting behavior.
 When $G=0$, the ground state is the closed-shell state with no decay width and
 the second pair-correlated state is a resonant state of the Fermion pair
 $S^\dagger_9$  with the half width $\Gamma/2=\gamma =1$.
  As the pairing correlations become stronger, the ``width'' of the ground state
 increases and after a critical strength of $G$ it exceeds the width of
 the second state which is originated in a resonant state.

   How does the imaginary part of the ground-state energy Im($E$) depend on
 the parameter $\gamma$?  The upper graph of Fig. 2 illustrates Im($E$)
 as a function of $\gamma/2$, in case of $N=8$ and $G=0.3$.  The value of Im($E$)
 is proportional to $\gamma$ in a small $\gamma$ region, but reaches the top
 at $\gamma/2 \approx 2$ after becoming a gentle slope around $\gamma/2 \sim 1$.
 To see the structure change, we calculated the occupation probabilities
 $v^2_\nu$ of respective s.p. states for the ground state.
 The real parts of $v^2_\nu$ for the resonant s.p. states are illustrated in the
 lower graph of Fig. 2.  This figure indicates that the occupation probabilities
 Re($v^2_\nu$) cross the zero point one after another. 
  Especially, Re($v^2_9$) of the lowest resonant s.p. level $\nu =9$ decreases
 drastically as $\gamma$ increases, and becomes negative at $\gamma/2 \approx 1.15$.
   The occupation probability must be positive to keep the physical meaning.
 If one of Re($v^2_\nu$) is negative, the state is unphysical.
 We should, therefore, stop our treatment in the unphysical situation
 for $\gamma/2 \gtrsim 1.15$.  This model calculation suggests
 that the resonant s.p. states with too large widths should be removed
 from the space of the pairing correlations.

\begin{figure}[h]
\includegraphics[width=6.8cm,height=6.7cm]{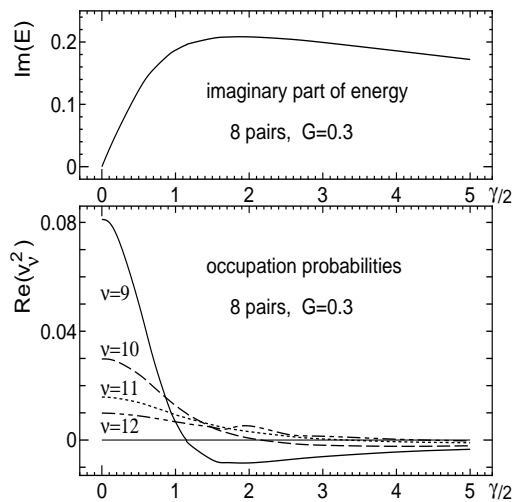}
  \caption{The imaginary part of the ground-state energy Im($E$) and the occupation
           probabilities Re($v^2_\nu$) of the resonant s.p. levels as a function
           of $\gamma/2$, in case of $N=8$ and $G=0.3$.}
  \label{fig2}
\end{figure}

   It should be noted that Figs. 1 and 2 give information about how the quantities
 vary when the pairing interactions are weakened for the resonant s.p. states
 or widths $\gamma$ are narrowed.

   When $N=9$, the last pair occupies the resonant s.p. levels in our model.
 This situation is similar to the system of a core plus one pair considered in Refs.
 \cite{Betan,Michel}, which discussed important roles of the resonant and continuum
 s.p. states in binding of the last one pair.  In Fig. 3, we show the total
 energy of the $N=9$ system (ground state) as a function of $G$.
 The real part Re($E$) is measured from the base level -64 that is the unperturbed core
 energy $\Sigma^8_{\nu =1}2(\nu -8.5)$.  If we regard the $N=8$ inner part as a core
 after Refs. \cite{Betan,Michel}, the system is reduced to one pair moving
 in the 4 resonant s.p. levels.
 Its energy shown by the dash curve (a) decreases slowly as $G$ increases and 
 becomes negative.  The negative energy means binding (confinement) of the last pair
 in the system, according to Ref. \cite{Michel}.  The imaginary part of energy Im($E$)
 remains being 1 in the truncated model (a) which has only 4 resonant s.p. levels
 with $\gamma/2=0.5$.
 
    If we take the 8 bound s.p. levels into account, the ground-state energy is
 very much lowered as shown by the solid curve (b) in Fig. 3.  The energy crosses
 the zero point at much smaller $G$ ($\approx 0.11$) as compared with (a).
 Obviously, the inclusion of the bound s.p. states strengthens the correlation energy.
  This, however, does not mean the binding of the last pair,
 because the 8 pair system without the last pair also gets large binding energy.
 The energy difference between the 9 and 8 pair systems, which is equal to the
 pair decay $Q$-value (we denote it by $Q_{\rm pair}$), is plotted by the dot curve
 (b$'$) in Fig. 3.  The curve indicates that the $N=9$ system cannot bind the last pair
 in our model assuming the same potential depth for $N=8$ and $N=9$.  We can say
 that the correlations of the last pair in the unbound s.p. states with the inner part
 are very important but whether the last pair is bound or not in the system may depend
 on complicated conditions. (It is expected that the $2N+1=17$ system is more loosely
 bound due to the blocking effect than the $2N=18$ system.)
  The imaginary part of energy Im($E$) increases also in the $N=9$ system as $G$
 increases, in our model without the continuum s.p. states.
 
\begin{figure}[h]
\includegraphics[width=6.8cm,height=6.5cm]{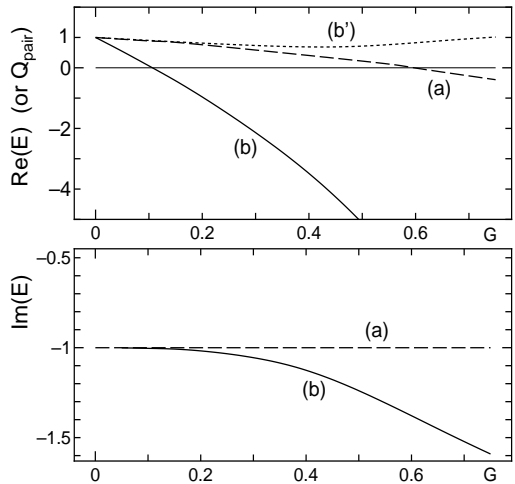}
  \caption{Ground-state energy of the $N=9$ system ($\gamma/2 =0.5$):
           (a) neglects the bound s.p. states; (b) includes them.}
  \label{fig3}
\end{figure}

   The previous work \cite{Betan,Michel} pointed out important roles of the
 continuum s.p. states in the core plus one pair system.
 The calculations \cite{Betan,Michel} showed that the continuum states contribute
 to the binding of the last pair in the system without bound s.p. states like (a). 
 If the continuum s.p. states are correctly treated, therefore, we can expect
 that the continuum states may contribute  to the binding of the last pair
 in the system including bound s.p. states like (b) and that the imaginary part
 of energy Im($E$) decreases reasonably.

  Now let us consider sixteen neutrons of $^{24}$O as an example of realistic
 system.  We neglect the continuum s.p. states, because we do not have
 a reliable treatment of them in our method.
  We generated the single-neutron basis for  $^{24}$O by a Woods-Saxon
 potential with the radius $R_0=1.2A^{1/3}$ fm,
 the surface diffuseness $a=0.65$ fm, the potential depth $V_0=-58$ MeV
 and the strength of the spin-orbit term $U_{so}=15$ MeV \cite{Vertse}. 
  The s.p. energies obtained are listed in the second column of Table I.
 The energies $\varepsilon_a$ are similar to those obtained with
 the Cwoik code \cite{Cwoik}, except for the very low energies of the $2p_{3/2}$ and
 $2p_{1/2}$ orbits.  The resonant widths are large for $2p_{3/2}$ and $2p_{1/2}$.
  Using the s.p. energies, we solved the Richardson equations for the lowest three
 pair-correlated states of $^{24}$O, by varying the pairing force strength $G$. 
 Though we omit figures, calculated result shows the increase of the pairing
 correlation energy as $G$ increases, which resembles the small $G$ region of Fig. 1.
 The general features shown by the model (\ref{eq:5}) are confirmed in $^{24}$O.

\begin{table}
\caption{Occupation probabilities $v^2_a$ of respective spherical orbits $a$
        in the ground state of $^{24}$O when $G=0.61$ MeV.}
\begin{tabular}{c|c|cc|cc|cc}   \hline
  & & \multicolumn{2}{c|}{neglect $\gamma_a$}
    & \multicolumn{2}{c|}{include $\gamma_a$}
     & \multicolumn{2}{c}{exclude $\circ$} \\
  & $\varepsilon_a$ (MeV) & $v^2_a$ & $\Delta_a$
    & Re($v^2_a$) & $\Delta_a$
     & Re($v^2_a$) & Re($v^2_a$) \\ \hline
 $1f_{7/2}$ & 5.49-i0.69 & 0.026 & 0.39 & 0.022 & 0.36 & 0.013   & $\circ$ \\
 $2p_{1/2}$ & 1.74-i3.06 & 0.069 & 0.15 & 0.018 & 0.08 & $\circ$ & $\circ$ \\
 $2p_{3/2}$ & 0.45-i2.06 & 0.110 & 0.38 & 0.047 & 0.26 & $\circ$ & $\circ$ \\
 $1d_{3/2}$ &  -0.07 & 0.137 & 0.42 & 0.137 & 0.42  & 0.082 & 0.056 \\ \hline
 $2s_{1/2}$ &  -3.12 & 0.759 & 0.26 & 0.835 & 0.23  & 0.923 & 0.960 \\
 $1d_{5/2}$ &  -4.54 & 0.872 & 0.61 & 0.908 & 0.53  & 0.960 & 0.978 \\
 $1p_{1/2}$ & -13.34 & 0.985 & 0.08 & 0.987 & 0.07  & 0.995 & 0.998 \\
 $1p_{3/2}$ & -16.26 & 0.989 & 0.11 & 0.991 & 0.11  & 0.997 & 0.999 \\
 $1s_{1/2}$ & -28.06 & 0.997 & 0.03 & 0.997 & 0.03  & 0.998 & 1.000 \\ \hline
 $\Delta$   &  & \multicolumn{2}{r|}{2.45}
               & \multicolumn{2}{r|}{2.09}
               & 1.27 & 0.73 \\
 $E$        &  & \multicolumn{2}{r|}{-189.95}
               & \multicolumn{2}{r|}{-189.29}
               & -187.77 & -186.91 \\ \hline
\end{tabular}
\label{table1}
\end{table}

 Table I lists the real parts of the occupation probabilities $v^2_a$
 and pairing gap $\Delta=\Sigma_a\Delta_a=\Sigma_aG(j_a+1/2)v_a\sqrt{1-v^2_a}$,
 for the ground state.
 Here, we set $G=0.61$ MeV so that $\Delta$ is nearly the standard value
 $12/\sqrt{A}$ MeV when $\gamma_a=0$.
 Table I indicates that the widths $\gamma_a$ disturb the pairing correlations
 and pairs hesitate in jumping to the resonant orbits with large widths.
 The pairing gap $\Delta$ is 2.45 MeV when neglecting $\gamma_a$, while it is
 2.09 MeV when we take account of $\gamma_a$.
  This feature agrees with that of the HF+BCS calculation \cite{Sand}.
 It is important to notice that the resonant orbits with large widths participate
 little in the pairing correlations.  This property is sharply revealed 
 if the widths of $2p_{3/2}$ and $2p_{1/2}$ are larger.
 When using another parameters for the Woods-Saxon potential, we had negative
 values of Re($v^2_a$) for $2p_{3/2}$ and $2p_{1/2}$, which corresponds to the
 unphysical situation pointed out in Fig. 2.
 
  Kruppa {\it et al.} \cite{Krup} argued that only resonant s.p. states with narrow
 widths should be included, to keep physical meaning.
 If we exclude the orbits $2p_{3/2}$ and $2p_{1/2}$ following them, the results
 become very different as shown in the seventh column of Table I.  Excluding
 $2p_{3/2}$ and $2p_{1/2}$ changes the occupation probabilities Re($v^2_a$) very much.
 The pairing gap $\Delta$ decreases from 2.09 MeV to 1.27 MeV.
 (It is interesting to note that the values of Re($v^2_a$) approach those of
  the fifth column including $2p_{3/2}$ and $2p_{1/2}$, if $G$ is enlarged
  so as to give the same gap $\Delta \approx 2.09$ MeV.)
 The significant reduction means that the resonant s.p. states with large widths,
 if they are included, are little occupied but still contribute
 to the pairing correlations.  If we neglect the last resonant s.p. orbit
 $1f_{7/2}$ further, the pairing correlations are much weakened as shown
 in the last column of Table I ($\Delta = 0.73$ MeV).
   On the ground-state energy tabulated in the bottom line of Table I,
 whether including or excluding the resonant s.p. states $2p_{3/2}$ and $2p_{1/2}$
 is more decisive than the effect of the resonant widths $\gamma_a$.
 Exclusion of $2p_{3/2}$ and $2p_{1/2}$ reduces the energy gain about 1.5 MeV
 (from 189.29 to 187.77), while the resonant widths $\gamma_a$ reduces 
 the energy gain 0.66 MeV (from 189.95 to 189.29).  The imaginary part of the
 ground-state energy Im($E$) becomes very small -0.09 MeV from -1.12 MeV,
 if the resonant orbits $2p_{3/2}$ and $2p_{1/2}$ with large widths are excluded,
 which complies with the physical demand \cite{Krup}.
 
   In conclusion, we have investigated the effects of the resonant s.p. states on the
 pairing correlations by an exact treatment of the pairing Hamiltonian.
 The present study shows that many-body correlations between bound and resonant
 particle pairs are important and that resonant s.p. states with too large widths
 should not be included in the space of the pairing correlations.
   We have not completed the study of the contribution from the continuum s.p. states.
 The interactions related to the resonant s.p. states and especially to the continuum
 s.p. states must be appropriately evaluated.
 Our approach using the Richardson equations is not adequate to this problem.
 It is necessary to use different pairing interactions in the next step.
 The conclusion that resonant s.p. states with very short life time
 should be excluded from the space of the pairing correlations
 must be investigated further in calculations which deal with
 the continuum s.p. states correctly.



\end{document}